# The enhancement of quantum entanglement under an open Dirac system with Hawking effect in Schwarzschild space-time

Wen-Yang Sun[1], Dong Wang[1,2], Bao-Long Fang[3], Jia-Dong Shi[1] and Liu Ye[1,*]

[1] *School of Physics & Material Science, Anhui University, Hefei 230601, People's Republic of China*

[2] *CAS Key Laboratory of Quantum Information, University of Science and Technology of China, Hefei 230026, People's Republic of China*

[3] *Department of Mathematics & Physics, Hefei University, Hefei 230601, People's Republic of China*

**Abstract:** In this letter, we mainly investigate how to enhance the damaged quantum entanglement under an open Dirac system with Hawking effect within Schwarzschild space-time. We consider that particle *A* hold by Alice undergoes generalized amplitude damping noise in a flat space-time and another particle *B* by Bob entangled with *A* is under a Schwarzschild space-time. Subsequently, we put forward a physical scheme to recover the damaged quantum entanglement by prior weak measurement on subsystem *A* before the interaction with the decoherence noise followed by post-measurement filtering operation. The results indicate that our scheme can effectively recover the damaged quantum entanglement affected by the Hawking effect and the noisy channel. Thus, our work might be beneficial to understand the dynamic behavior of quantum state and recover the damaged quantum entanglement with open Dirac systems under Hawking effect in the background of Schwarzschild black hole.

## 1. Introduction

Quantum entanglement has been a topic of great interest ever since the pioneering work presented by Einstein *et al.* [1]. It is the central concept within quantum information theory. Entanglement can be defined as the inseparability of quantum states [2-7]. It has been viewed as an important resource within quantum information processing tasks, such as quantum teleportation [8-10], quantum key distribution [11], quantum cryptograph [12, 13], and quantum remote preparation [14-19]. Recently, more and more efforts have been made on the research of quantum entanglement under the relativistic framework by discussing how the Unruh-Hawking effect influences the degree of entanglement [20-27], because it plays a vital role during achieving various quantum information processing tasks. Undoubtedly, the research of entanglement dynamics behaviors in a relativistic setting will not only provide a more complete framework for the quantum information theory, but also take a key part in the understanding of the entropy and information paradox of black holes [28].

[*] Corresponding author: yeliu@ahu.edu.cn

However, most of the current investigations are restricted to the investigation of quantum information in an isolated quantum system. Actually, a realistic quantum system unavoidably suffers from external environment, and leads to the decoherence. Decoherence effect has been studied in the cavity QED and other experiments [29, 30], and it plays a fundamental role in the description of the quantum-to-classical transition [31, 32]. This bidirectional interaction between the system and its external noisy environment will lead to the degradation of quantum correlation or coherence. In certain cases, it will give rise to quantum entanglement sudden death (QESD). There exists a fact that it is important to restore the damaged entanglement under an open Dirac system with Hawking effect within Schwarzschild space-time. Hence, we naturally arise a question: how to enhance or recover the damaged quantum entanglement for an open Dirac system within the curved space-time?

To solve the problem, we will investigate how to recover the lost quantum entanglement for an open Dirac system with the effect of Hawking radiation [33, 34]. Our aim is to focus on enhancing quantum entanglement with Hawking effect under a noisy channel, which may lead to much better enhance entanglement with Hawking effect in the presence of decoherence noise for quantum information processing. Our work contributes to exploring the situation that there are two observers, Alice and Bob, sharing an initial entangled state at the same initial point in the flat Minkowski space-time. Then, Bob freely falls in toward Schwarzschild black hole and finally locates near the event horizon, and Alice suffers from the decoherence noise. Besides, as a noise model for decoherence [35, 36], we chose generalized amplitude damping (GAD) channel [37], a reservoir in thermal equilibrium with a single qubit at a finite temperature. The GAD channel can be represented in terms of Kraus operators

$$P_1 = \sqrt{h}\begin{pmatrix} 1 & 0 \\ 0 & \sqrt{1-e} \end{pmatrix}, \quad P_2 = \sqrt{h}\begin{pmatrix} 0 & \sqrt{e} \\ 0 & 0 \end{pmatrix},$$
$$P_3 = \sqrt{1-h}\begin{pmatrix} \sqrt{1-e} & 0 \\ 0 & 1 \end{pmatrix}, P_4 = \sqrt{1-h}\begin{pmatrix} 0 & 0 \\ \sqrt{e} & 0 \end{pmatrix},$$
(1)

where the parameter $e$ can be expressed as $[1-\exp(-\gamma t)]$ through the coupling constant $\gamma$ (which defines the temperature of the reservoir, for instance) and the time of interaction $t$. Here, $\{\gamma, h\}$ is usually a function of environment temperature $T'$.

$$\gamma = \gamma' \left[ 1 + \frac{2}{\exp(-\hbar\omega/k_B T')-1} \right],$$
$$h = \frac{1}{\exp(-\hbar\omega/k_B T')+1}, \qquad (2)$$

where $\gamma'$ is the energy relaxation rate, $\hbar\omega$ is the transition energy of quantum system and $k_B$ is the Boltzmann constant. Note that setting $h=1$ would reduce GAD channel to the well-informed amplitude damping (AD) channel. The dissipation course under GAD channel can be described by a quantum operation as [3]

$$\varepsilon_{GAD}(\rho) = \sum_{i=1}^{4} \left( P_i^A \otimes I^B \right) \rho \left( P_i^A \otimes I^B \right)^\dagger, \qquad (3)$$

Alice undergoes the GAD channel. How to finish the aim of enhancing quantum entanglement, we give a detailed physical scheme shown in Fig. 1.

This letter is organized as follows: In Sec. 2, the Hawking radiation for Dirac fields in the Schwarzschild space-time is reviewed briefly. In Sec. 3, a physical scheme of enhancing quantum entanglement is put forward. Finally, we summarize our work in Sec. 4.

## 2. Hawking radiation for Dirac fields in Schwarzschild space-time

In this section, we introduce a metric Hawking radiation for Dirac fields under the Schwarzschild space-time, it can be written as

$$ds^2 = -(1-\frac{2M}{r})dt^2 + (1-\frac{2M}{r})^{-1}dr^2 + r^2(\sin^2\theta d\varphi^2 + d\theta^2), \qquad (4)$$

where the parameter $M$ represents the mass of the black hole. For simplicity, we consider $G$, $c$, $\hbar$ and $k_B$ as unity here. For the Schwarzschild space-time, the Dirac equation [38] $\psi[\gamma^a e_a^u(\partial_u + \Gamma_u)] = 0$ under a curved space-time can be expressed as

$$-\frac{\gamma_0}{\sqrt{1-\frac{2M}{r}}}\frac{\partial \psi}{\partial t} + \gamma_1\sqrt{1-\frac{2M}{r}}[\frac{\partial}{\partial r} + \frac{1}{r} + \frac{M}{2r(r-2M)}]\psi + \frac{\gamma_2}{r}(\frac{\partial}{\partial \theta} + \frac{\cot\theta}{2})\psi + \frac{\gamma_3}{r\sin\theta}\frac{\partial \psi}{\partial \phi} = 0. \quad (5)$$

On the one hand, we can obtain the positive (fermions) frequency outgoing solutions for the outside region Ⅰ and inside region Ⅱ of the event horizon [39] by solving Eq. (5)

$$\psi_k^{\text{I}+} = \xi e^{-i\omega u} \ (r > r_+),$$
$$\psi_k^{\text{II}+} = \xi e^{i\omega u} \ (r < r_+). \tag{6}$$

where $\xi$ is a 4-component Dirac spinor [40], $\omega$ is a monochromatic frequency of Dirac flied, $u = t - r^*$ and $r^* = \left(2M \ln[(r-2M)/(2M)] + r\right)$ represent the tortoise coordinate. Particles and antiparticles will be classified with respect to the future directed time-like Killing vector in each region. Besides, the generalized light-like Kruskal coordinates under the Schwarzschild space-time are introduced as follows

$$u = -4M \ln\left(\frac{U}{4M}\right), \ v = 4M \ln\left(\frac{V}{4M}\right), \ if(r < r_+),$$
$$u = -4M \ln\left(\frac{-U}{4M}\right), \ v = 4M \ln\left(\frac{V}{4M}\right), \ if(r > r_+), \tag{7}$$

and based on the Damour-Ruffini's suggestion [41], where $U$ and $V$ are the generalized light-like Kruskal coordinates under the Schwarzschild space-time, and regularly across the past and future horizons of the extended space-time [40]. One finds another complete basis for positive energy modes by making an analytic continuation for Eq. (6)

$$\vartheta_k^{\text{I}+} = e^{2\pi \cdot \omega_k M} \psi_k^{\text{I}+} + e^{-2\pi \cdot \omega_k M} \psi_{-k}^{\text{II}-},$$
$$\vartheta_k^{\text{II}+} = e^{2\pi \cdot \omega_k M} \psi_k^{\text{II}+} + e^{-2\pi \cdot \omega_k M} \psi_{-k}^{\text{I}-}, \tag{8}$$

where $\omega_k$ is the frequency of the particles for the mode $k$. Then, the Bogoliubov transformations [42] between the annihilation operator and creation operator in the Schwarzschild and Kruskal coordinates can be given through quantizing the Dirac fields in the Schwarzschild and Kruskal modes, respectively. Through calculations, after properly normalizing the state vector, the ground and excited states of the Kruskal particle for mode $k$ can be expressed as

$$|0\rangle_k^+ \to \mu |0_k\rangle_{\text{I}}^+ |0_{-k}\rangle_{\text{II}}^- + \upsilon |1_k\rangle_{\text{I}}^+ |1_{-k}\rangle_{\text{II}}^-,$$
$$|1\rangle_k^+ \to |1_k\rangle_{\text{I}}^+ |0_{-k}\rangle_{\text{II}}^-, \tag{9}$$

where $\mu = [1+\exp(-\omega_k/T)]^{-\frac{1}{2}}$, $\upsilon = [1+\exp(\omega_k/T)]^{-\frac{1}{2}}$, and $T$ is the Hawking temperature [40] with $T = \frac{1}{8\pi M}$, $|n_{-k}\rangle_{\text{II}}^-$ and $|n_k\rangle_{\text{I}}^+$ correspond to the orthonormal bases for the inside and outside regions of the event horizon, respectively. For simplicity, $|n_k\rangle_{\text{I}}^+$ and $|n_{-k}\rangle_{\text{II}}^-$ are replaced by $|n\rangle_{\mathrm{I}}$ and $|n\rangle_{\mathrm{II}}$, respectively.

## 3. Physical scheme for enhancing quantum entanglement

Now, let us introduce two local operations (weak measurement and filtering operation), respectively. We take a brief introduce about weak measurement [43-46]. In practice, weak measurement can be realized by a device that monitors a qubit indirectly. If the device signals, which means the qubit transition $|1\rangle \rightarrow |0\rangle$, then we discard the result. If the device has no signal, that is, null outcome, the qubit state was only partially collapsed and we let it evolve. The local non-unitary operator (weak measurement) for a single qubit can be written as

$$M_{wk} = \begin{pmatrix} 1 & 0 \\ 0 & \sqrt{1-m} \end{pmatrix}, \tag{10}$$

where $m$ is the strength of weak measurement.

Besides, after one-qubit suffers from noisy environment, we can perform a filtering operation on a single qubit. This operation for a single qubit can be given by [47]

$$M_{ft} = \begin{pmatrix} \sqrt{1-f} & 0 \\ 0 & \sqrt{f} \end{pmatrix}, \qquad 0 < f < 1 \tag{11}$$

where $f$ is the strength of filtering operation. Filtering is a non-trace-preserving map which is known to be capable of increasing entanglement with some probability [47, 48]. In fact, this map can be realized as a null-result weak measurement [49].

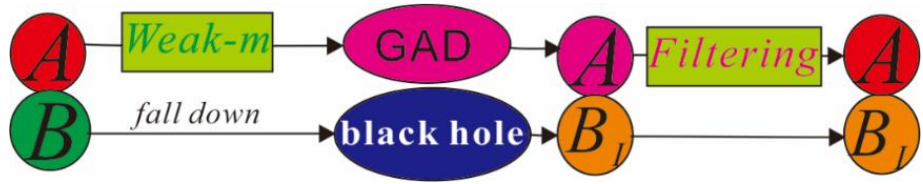

**Fig.1** The physical scheme diagram: Alice and Bob, sharing an initial entangled state at the same initial point in the flat Minkowski space-time. Then, Bob freely falls in toward Schwarzschild black hole and finally locates near the event horizon. Besides, Alice makes a weak measurement operation and then suffers from GAD noise, subsequently, we perform a filtering operation on Alice.

It is well known that the degree of entanglement for two-qubit system can be quantified conveniently by concurrence. Hence, we chose concurrence as entanglement measure. Generally speaking, concurrence can be defined as [50-52]

$$C(\rho) = \max\left\{0,\ \sqrt{\lambda_1} - \sqrt{\lambda_2} - \sqrt{\lambda_3} - \sqrt{\lambda_4}\right\},\quad \lambda_1 \geq \lambda_2 \geq \lambda_3 \geq \lambda_4 \geq 0, \tag{12}$$

where $\lambda_i (i=1, 2, 3, 4)$ are the eigenvalues of the matrix $R = \rho(\sigma_y \otimes \sigma_y)\rho^*(\sigma_y \otimes \sigma_y)$. If the density matrix is an $X$-structure, there is a reduced form for concurrence shown as

$$C(\rho^X) = \max\left\{0,\ 2\left(\sqrt{\rho_{14}\rho_{41}} - \sqrt{\rho_{22}\rho_{33}}\right),\ 2\left(\sqrt{\rho_{23}\rho_{32}} - \sqrt{\rho_{11}\rho_{44}}\right)\right\}, \tag{13}$$

where $\rho_{ij}$ are the elements of the matrix $\rho^X$. Suppose that Alice has a detector which is sensitive only to mode $|n\rangle_A$ and Bob has a detector which is sensitive only to mode $|n\rangle_B$, and they share an initial entangled state at the same initial point in the flat Minkowski space-time

$$|\psi\rangle_{AB} = \alpha|00\rangle_{AB} + \sqrt{1-\alpha^2}|11\rangle_{AB},\quad 0 < \alpha < 1. \tag{14}$$

Then, Bob freely falls in toward Schwarzschild black hole and finally locates near the event horizon. Via Eq. (9), Alice and Bob share the quantum state in the outside regions I of the event horizon, which can be given by

$$\begin{aligned}\rho_I^{AB} =\ & \alpha^2\mu^2|00\rangle\langle 00| + \alpha\mu\sqrt{1-\alpha^2}|00\rangle\langle 11| + \alpha\mu\sqrt{1-\alpha^2}|11\rangle\langle 00| \\ & + \alpha^2\upsilon^2|01\rangle\langle 01| + (1-\alpha^2)|11\rangle\langle 11|.\end{aligned} \tag{15}$$

Besides, Alice makes a weak measurement operation and then suffers from GAD noisy environment, subsequently, we perform a filtering operation on Alice. By utilizing Eqs. (1), (10), (11) and (15), we can obtain the final quantum state $\rho_{AB_1}$ as follows

$$\rho_{AB_1} = \frac{1}{Z}\begin{pmatrix} \rho_{11} & 0 & 0 & \rho_{14} \\ 0 & \rho_{22} & 0 & 0 \\ 0 & 0 & 0 & 0 \\ \rho_{41} & 0 & 0 & \rho_{44} \end{pmatrix}, \tag{16}$$

with

$$\begin{aligned}\rho_{11} &= \alpha^2\mu^2(1-f)(1+e(h-1)), \\ \rho_{22} &= (f-1)\left(\alpha^2(e-1-eh)\upsilon^2 - (\alpha^2-1)eh(m-1)\right), \\ \rho_{44} &= (1-\alpha^2)f(e(-1+2h)-1)(m-1), \\ \rho_{14} &= \rho_{41} = \alpha\mu\sqrt{1-\alpha^2}\sqrt{1-e}\sqrt{1-m}\sqrt{f(1-f)}, \\ Z &= \alpha^2\mu^2(1-f)(1+e(h-1)) + (1-\alpha^2)f(e(-1+2h)-1)(m-1) \\ & \quad + (f-1)\left(\alpha^2(e-1-eh)\upsilon^2 - (\alpha^2-1)eh(m-1)\right).\end{aligned}$$

The detailed calculation procedures are

$$\rho_1^{AB'} = \frac{1}{Z_1}\left((M_{wk} \otimes \mathbb{I}_2) \cdot \rho_1^{AB} \cdot (M_{wk} \otimes \mathbb{I}_2)\right), \tag{17-a}$$

$$\rho_1^{AB''} = \frac{1}{Z_1}\left(\sum_{i=1}^{4}(P_i \otimes \mathbb{I}_2) \cdot \rho_1^{AB'} \cdot (P_i \otimes \mathbb{I}_2)^\dagger\right), \tag{17-b}$$

$$\rho_{AB_1} = \frac{1}{Z}\left((M_{ft} \otimes \mathbb{I}_2) \cdot \rho_1^{AB''} \cdot (M_{ft} \otimes \mathbb{I}_2)\right), \tag{17-c}$$

where $\mathbb{I}_2$ is a 2×2 identity matrix. For clarity, the physical scheme sketch of the total system is depicted in Fig. 1. From Eq. (13), we can obtain the concurrence $C_{AB_1}$ of the state $\rho_{AB_1}$ as follow

$$C_{AB_1} = \max\left\{0, \frac{2\alpha\mu\sqrt{1-\alpha^2}\sqrt{1-e}\sqrt{1-m}\sqrt{f(1-f)}}{Z}\right\}. \tag{18}$$

Now we have two control parameters $m$ and $f$ at hand, which can manipulate the qubits' entanglement for a certain purpose at any time during the evolution. If we would like to get the best effect of the entanglement recovery, one needs to make optimization regarding filtering operation strength, the optimal filtering operation strength $f_0$ that maximizes the concurrence can be determined from the conditions $\partial C_{AB_1}/\partial f = 0$ and $\partial^2 C_{AB_1}/\partial^2 f < 0$. Hence, one can obtain that the optimal filtering operation strength is

$$f_0 = \frac{e\left(\alpha^2\left(\mu^2+\upsilon^2\right)+h\left(m-1-\alpha^2\left(m-1+\mu^2+\upsilon^2\right)\right)\right)-\alpha^2\left(\mu^2+\upsilon^2\right)}{\alpha^2\left(1-m-\mu^2-\upsilon^2+e(h-1)(m-1-\mu^2-\upsilon^2)\right)-(e(h-1)-1)(m-1)}. \tag{19}$$

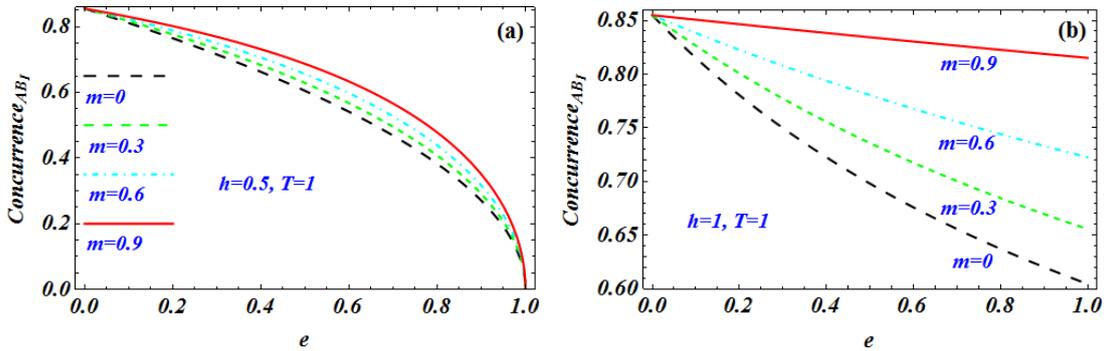

**Fig.2** (a), (b) The concurrence $C_{AB_1}$ as a function of decoherence strength $e$ for the different $m$ under the condition of $h=0.5$, $T=1$, $\alpha=\sqrt{2}/2$, $\omega_k=1$ and $h=1$, $T=1$, $\alpha=\sqrt{2}/2$, $\omega_k=1$, respectively

Now, let us discuss the relationship between the entanglement (concurrence) and the decoherence strength $e$ for the different weak measurement strength $m$ in Fig. 2. Herein, we utilize the optimal filtering operation strength. One can see that entanglement always decreases with the increase of decoherence strength $e$, and QESD always appears in GAD channel. However, the QESD will never happen when $h=1$ (Here, the noisy channel is an AD channel) and the weak measurement strength is a nonzero value. Certainly, quantum entanglement always increases with the increase of weak measurement strength in the presence of Hawking radiation and decoherence noise. It turns out that we obtain a better effect for entanglement recovery. As well as, our scheme can effectively recover the spoiled entanglement.

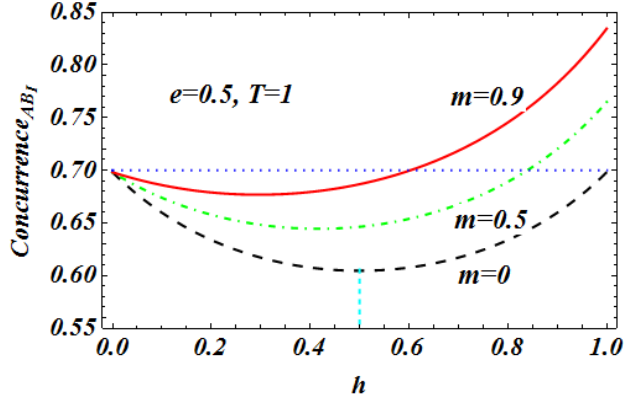

**Fig.3** The concurrence versus $h$ under the different $m$ with $e=0.5$, $T=1$, $\alpha=\sqrt{2}/2$, $\omega_k=1$.

Next, we illuminate the dynamics of the entanglement versus the parameter $h$ for the different weak measurement strength with $e=0.5$, $T=1$, $\alpha=\sqrt{2}/2$, $\omega_k=1$ in Fig. 3. The results show that quantum entanglement decreases at the beginning and then increases with the increase of the parameter $h$, no matter what the strength of weak measurement is. Additionally, it is interesting that the concurrence of $\rho_{AB_1}$ is symmetrical with $m=0$. However, when weak measurement strength is nonzero, the symmetry is destroyed. Subsequently, we discuss how the Hawking temperature affects the dynamics of quantum entanglement for different weak measurement strengths. As shown in Fig. 4, one can obtain that concurrence always decreases with the increase of Hawking temperature in the presence of decoherence noise, but QESD never occurs. It effectively manifest that the Hawking effect and decoherence noise can destroy quantum entanglement. However, the collective effect of weak measurement and filtering operation can recover the damaged quantum entanglement. In addition, due to the Pauli Exclusion Principle in Fermi-Dirac statistics, the quantum states cannot be excited infinitely with Hawking effect for Fermi particles. This readily explains why the physically accessible quantum entanglement does not decrease to absolutely vanish, though hypothesis Hawking temperature increases to infinity. Furthermore, as shown in Fig. 5, these local operations can obviously recover the spoiled quantum entanglement affected by the Hawking effect and decoherence as well. Thus, our scheme can gain a better preservation for quantum entanglement when Hawking temperature is relatively smaller.

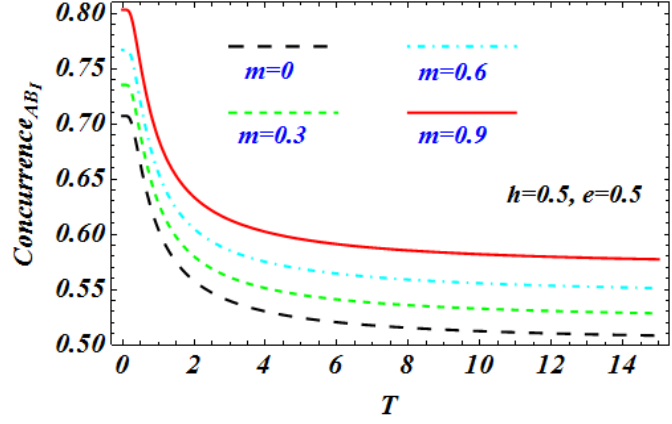

**Fig.4** The concurrence as a function of Hawking temperature $T$ for the different $m$ under the condition of $\alpha = \sqrt{2}/2$, $e = 0.5$, $h = 0.5$, $\omega_k = 1$.

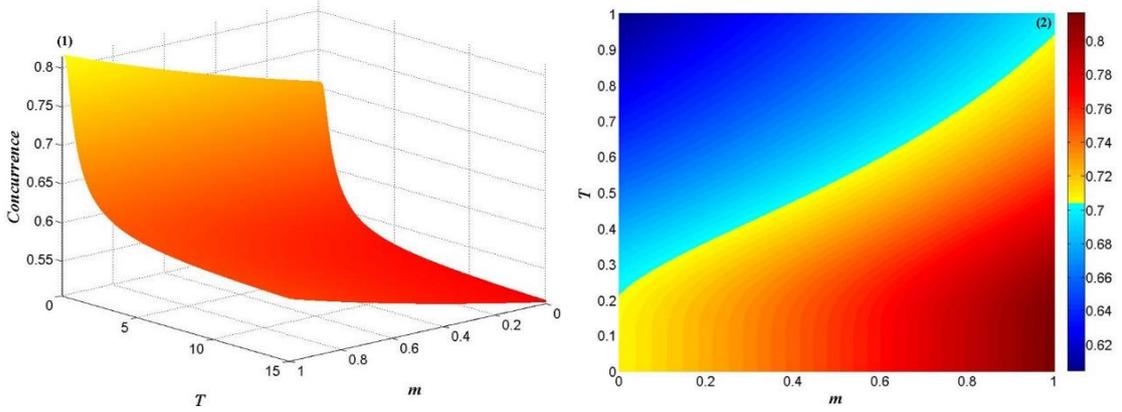

**Fig.5 (1)** Quantum entanglement as a function of Hawking temperature $T$ and weak measurement strength $m$ with $e = 0.5$, $h = 0.5$, $\alpha = \sqrt{2}/2$, $\omega_k = 1$, and **(2)** contour plot of concurrence versus weak measurement strength $m$ and Hawking temperature $T$ with $e = 0.5$, $h = 0.5$, $\alpha = \sqrt{2}/2$, $\omega_k = 1$.

## 4. Conclusions

To conclude, we examine how to recover the damaged quantum entanglement under an open Dirac system with Hawking effect in the background of Schwarzschild black hole. We have put forward a physical scheme to enhance the damaged quantum entanglement. For GAD channel, the results indicate that quantum entanglement always decreases with the increase of decoherence strength, and QESD always occurs. Nevertheless, QESD never happens when $h=1$ (i.e., the noisy channel reduces to an AD channel), the weak measurement strength is a nonzero value and the optimal filtering operation is implemented. Besides, quantum entanglement decreases at the beginning and then increases with the increase of the parameter $h$ in the presence of Hawking radiation, no matter what the weak measurement strength is. As well as, one can also obtain that quantum entanglement always decreases with the increase of Hawking temperature in the presence of decoherence noise, and QESD never appears. Furthermore, quantum entanglement always increases with the increasing weak measurement strength in the presence of Hawking radiation and decoherence noise, when we implement the optimal filtering operation. These results testify that the collective effect of weak measurement and filtering operation can effectively recover the damaged quantum entanglement, although the Hawking effect and decoherence noise can destroy the original quantum entanglement.

## Acknowledgments

This work was supported by the National Science Foundation of China under Grant Nos. 11575001 and 61601002, Anhui Provincial Natural Science Foundation (Grant No. 1508085QF139) and Natural Science Foundation of Education Department of Anhui Province (Grant No. KJ2016SD49), and also the fund from CAS Key Laboratory of Quantum Information (Grant No. KQI201701).

## References

[1] A. Einstein, B. Podolsky and N. Rosen, Phys. Rev. **47**, (1935) 777.

[2] R. Horodecki, P. Horodecki, M. Horodecki, and K. Horodecki, Rev. Mod. Phys. **81**, (2009) 865.

[3] M.A. Nilsen and I.L. Chuang, Quantum Computation and Quantum Communication. (Cambridge University Press, Cambridge, 2000).

[4] C.H. Bennett and D.P. DiVincenzo, Nature (London) **404**, (2000) 247.

[5] J.S. Bell, Physics (Long Island City, N.Y.) **1**, (1964) 195.

[6] S.B. Zheng and G.C. Guo, Phys. Rev. Lett. **85**, (2000) 2392.

[7] W.-Y. Sun, D. Wang, B.-L. Fang, L. Ye, Laser Phys. Lett. **15**, (2018) 035203.

[8] V.A. Guillaume, M.M. Eduardo, K. Achim, Phys. Rev. A **93**, (2016) 022308.

[9] Z. Zhao, Y.A. Chen, A.N. Zhang, T. Yang, H.J. Briegel, J.W. Pan, Nature **430**, (2004) 54.

[10] L.Y. Hu, Z.Y. Liao, S.L. Ma, and M. Suhail Zubairy, Phys. Rev. A **93**, (2016) 033807.

[11] X.B. Wang, Phys. Rev. A **87**, (2013) 012320; Q. Wang, X.B. Wang, Sci. Rep. **4**, (2014) 4612; Z.W. Yu, Y.H. Zhou, X.B. Wang, Phys. Rev. A **91**, (2015) 032318.

[12] C. Ottaviani, S. Pirandola, Sci. Rep. **6**, (2016) 22225.

[13] S.N. Molotkov and T.A. Potapova, Laser Phys. Lett. **13**, (2016) 035201.

[14] H.K. Lo, Phys. Rev. A **62**, (2000) 012313.

[15] A.K. Pati, Phys. Rev. A **63**, (2000) 014302.

[16] C.H. Bennett, D.P. DiVincenzo, P.W. Shor, J.A. Smolin, B.M. Terhal, and W.K. Wootters, Phys. Rev. Lett. **87**, (2001) 077902.

[17] H.Q. Liang, J.M. Liu, S.S. Feng, J.G. Chen, and X.Y. Xu, Quantum Inf. Process. **14**, (2015) 3857.

[18] D. Wang, Y.D. Hu, Z.Q. Wang, L. Ye, Quantum Inf. Process. **14**, (2015) 2135.

[19] D. Wang, R.D. Hoehn, L. Ye and S. Kais, Entropy **17**, (2015) 1755.

[20] G. Adesso, I. Fuentes-Schuller, and M. Ericsson, Phys. Rev. A **76**, (2007) 062112.

[21] M. Aspachs, G. Adesso, and I. Fuentes, Phys. Rev. Lett. **105**, (2010) 151301.

[22] L. Lamata, M.A. Martin-Delgado, and E. Solano, Phys. Rev. Lett. **97**, (2006) 250502; O. Viyuela, A. Rivas, and M. A. Martin-Delgado, Phys. Rev. Lett. **112**, (2014) 130401.

[23] J. León and E. Martín-Martínez, Phys. Rev. A **80**, (2009) 012314.

[24] Q.Y. Pan and J.L. Jing, Phys. Rev. D **78**, (2008) 065015.

[25] J.C. Wang and J.L. Jing, Phys. Rev. A **83**, (2011) 022314.

[26] Y. Ling, S. He, W.G. Qiu, and H.B. Zhang, J. Phys. A **40**, (2007) 9025.

[27] I. Fuentes-Schuller and R.B. Mann, Phys. Rev. Lett. **95**, (2005) 120404.

[28] S. Xu, X.-K. Song, J.-D. Shi, and L. Ye, Phys. Rev. D **89**, (2014) 065022.

[29] M. Brune, E. Hagley, J. Dreyer, X. Maitre, A. Maali, C. Wunderlich, J.M. Raimond, S. Haroche, Phys. Rev. Lett. **77**, (1996) 4887.

[30] C.J. Myatt, B.E. King, Q.A. Turchette, C.A. Sackett, D. Kielpinski, W.M. Itano, C. Monroe, D.J. Wineland, Nature **403**, (2000) 269.

[31] D. Giulini, E. Joos, C. Kiefer, J. Kupsch, I.O. Stamatescu, and H.D. Zeh, Decoherence and the Appearance of a Classical World in Quantum Theory (Springer, 1996).

[32] M.A. Schlosshauer, Decoherence and the Quantum-to-Classical Transition (Springer, 2007).

[33] S.W. Hawking, Nature **248**, (1974) 30.

[34] S.W. Hawking, Phys. Rev. D **14**, (1976) 2460.

[35] W.-Y. Sun, D. Wang, J.-D. Shi, L. Ye, Sci. Rep. **7**, (2017) 39651.

[36] W.-Y. Sun, D. Wang, J. Yang, L. Ye, Quantum Inf. Process **16**, (2017) 90.

[37] S.C. Wang, Z.W. Yu, W.J. Zou, X.B. Wang, Phys. Rev. A **89**, (2014) 022318.

[38] D.R. Brill and J.A. Wheeler, Rev. Mod. Phys. **29**, (1957) 465.

[39] J.L. Jing, Phys. Rev. D **70**, (2004) 065004.

[40] J.C. Wang, Q.Y. Pang, and J.L. Jing, Ann. Phys. **325**, (2010) 1190-1197.

[41] T. Damoar and R. Ruffini, Phys. Rev. D **14**, (1976) 332.

[42] S.M. Barnett and P.M. Radmore, Methods in Theoretical Quantum Optics, Oxford University Press, New York, (1997) pp.67-80.

[43] A.N. Korotkov, A.N. Jordan, Phys. Rev. Lett. 97, 166805 (2006); A.N. Korotkov, K. Keane, Phys. Rev. A **81**, (2010) 040103.

[44] Q. Sun, M. Al-Amri, and M. Suhail Zubairy, Phys. Rev. A **80**, (2009) 033838.

[45] J.C. Lee, Y.C. Jeong, Y.S. Kim, Y.H. Kim, Opt. Express **19**, (2011) 16309.

[46] J.S. Xu, K. Sun, C.F. Li, X.Y. Xu, G.C. Guo, E. Andersson, R. Lo Franco, G. Compagno, Nat. Commun. **4**, (2013) 2851.

[47] M. Siomau and A. A. Kamli, Phys. Rev. A **86**, 032304 (2012).

[48] M. Tiersh, Ph.D. thesis, University of Freiburg, (2009).

[49] Q. Sun, M. Al-Amri, L. Davidovich, and M.S. Zubairy, Phys. Rev. A **82**, (2010) 052323.

[50] W.K. Wootters, Phys. Rev. Lett. **80**, (1998) 2245.

[51] W.-Y. Sun, D. Wang, Z.-Y. Ding, L. Ye, Laser Phys. Lett. **14**, (2017) 125204.

[52] V. Coffman, J. Kundu, W.K. Wootters, Phys. Rev. A **61**, (2000) 052306.